\documentstyle[12pt]{article}

\newcommand {\nn}    {\nonumber}
\newcommand {\vs}[1]  { \vspace*{#1 cm} }
\newcounter{eq}
\newcounter{sc}

\newcommand {\PL}   {Phys.Lett.}
\newcommand {\PR}   {Phys.Rev.}
\newcommand {\PRL}   {Phys.Rev.Lett.}
\newcommand {\CMP}  {Comm.Math.Phys.}

\def\overleftrightarrow#1{\vbox{\ialign{##\crcr
 $\leftrightarrow$\crcr\noalign{\kern-1pt\nointerlineskip}
 $\hfil\displaystyle{#1}\hfil$\crcr}}}

\newlength{\minitwocolumn}
\begin{document}

%%%%%%%%%%%%%%%%%%%%%%%%%%%%%%%%%%%%%%%%%%%%%%%%%%%%%%%%%%%%%%%%%%
%%%%%%%%%%%%%%%%%%%%%%%% Title %%%%%%%%%%%%%%%%%%%%%%%%%%%%%%%%%%%
%%%%%%%%%%%%%%%%%%%%%%%%%%%%%%%%%%%%%%%%%%%%%%%%%%%%%%%%%%%%%%%%%%

\begin{flushright}
EDO-EP-32\\
August, 2000\\
\end{flushright}
\vspace{30pt}

%\magnification=\magstep1
\pagestyle{empty}
\baselineskip15pt
%\font\cmssB=cmss17
%\font\cmssS=cmss10

\begin{center}
{\large\bf Reissner-Nordstrom Black Hole in Gravity
Localized Models

 \vskip 1mm
}

\vspace{20mm}

Ichiro Oda
          \footnote{
          E-mail address:\ ioda@edogawa-u.ac.jp
                  }
\\
\vspace{10mm}
          Edogawa University,
          474 Komaki, Nagareyama City, Chiba 270-0198, JAPAN \\

\end{center}

%\maketitle

\vspace{15mm}
\begin{abstract}
We investigate the possibility of having the electrically charged 
Reissner-Nordstrom black hole in the gravity localized models in a
brane world. 
It is shown that the Reissner-Nordstrom black hole
exists as a solution in the 5D Randall-Sundrum domain wall model 
if there is the $U(1)$ bulk gauge field. We find that the charged 
black hole is localized on a 3-brane even if zero mode of the bulk 
gauge field is not in general localized on the brane in the domain 
wall model. 
Moreover, we extend these observations to the higher-dimensional 
topological defect models with codimension more than one in a general 
space-time dimension such as the string-like defect model with codimension 2 
in six dimensions and the monopole-like defect model with codimension 3 
in seven dimensions e.t.c.     

\vspace{15mm}

\end{abstract}

\newpage
\pagestyle{plain}
\pagenumbering{arabic}
%\setcounter{page}{1}

%%%%%%%%%%%%%%%%%%%%%%%%%%%%%%%%%%%%%%%%%%%%%%%%%%%%%%%%%%%%%%%%%%
%%%%%%%%%%%%%%%%%%%%%%%% Article %%%%%%%%%%%%%%%%%%%%%%%%%%%%%%%%%
%%%%%%%%%%%%%%%%%%%%%%%%%%%%%%%%%%%%%%%%%%%%%%%%%%%%%%%%%%%%%%%%%%

\rm
%%%%%%%%%%%%%%%%%%%%%%%%%%%%%%%%%%%%%%%%%%%%%%%%%%%%%%%%%%%%%%%%%%%%%
%%%%%%%%%%%%%%%%%%%%%%%%%%%%%%   SEC  1    %%%%%%%%%%%%%%%%%%%%%%%%%%
%%%%%%%%%%%%%%%%%%%%%%%%%%%%%%%%%%%%%%%%%%%%%%%%%%%%%%%%%%%%%%%%%%%%%
\section{Introduction}

Black holes have so far played a crucial role in the development of
quantum gravity and superstring theory. For instance, the discovery
of the Hawking radiation \cite{Hawking} that black holes are not, 
after all, completely black but emit thermal radiation with a definite 
temperature due to quantum effects has triggered the widespread interest 
in quantum field theory in curved space-time. And the recent progress of a
precise calculation of the black hole entropy and string dualities 
in superstring theory suggests that black holes do obey the ordinary rules 
of quantum mechanics so the laws of quantum mechanics do not need to
be modified radically down to the Planck scale \cite{Pol}. 
Thus the study of black holes would give us the most valuable clues of 
constructing a quantum theory of gravity and a unified theory of 
fundamental interactions in future.

In recent years, we have witnessed a vast interest in the idea of a
brane world where our world is considered to be a 3-brane embedded
in a higher dimensional space-time \cite{Rubakov, Akama, Visser,
Arkani-Hamed, Randall2, Oda1, Oda2}. 
The key point of this scenario is localization of gauge and matter 
fields on the world-volume of the 3-brane, for which D-branes 
\cite{Polchinski} would provide a natural framework.   
Another key point is that even gravity, which watches the whole
space-time structure so lives in the whole space-time,
can be also localized on the 3-brane in the sense that a normalizable
graviton zero mode trapped by the brane reproduces four dimensional
Newton's law with only a small correction \cite{Randall2}.

It is then natural to ask whether or not there are black hole solutions
on a 3-brane in a well-known gravity localized model in $AdS_5$, the 
Randall-Sundrum model \cite{Randall2} in a brane world. 
Actually, in the model, since it turns out that the geometry on the brane 
is Ricci-flat, one can find the Schwarzschild geometry as a solution 
to Einstein's equations \cite{Chamblin}. 
On the other hand, the Reissner-Nordstrom black 
hole satisfies Einstein's equations with the stress-energy of the electric
field as a source, so it appears to be difficult to have the charged black
hole in the Randall-Sundrum model. Related to this point, it is recently 
shown that it is impossible to embed the Reissner-Nordstrom black hole in the 
gravity localized model via the conventional Kaluza-Klein reduction of the 
bulk metric \cite{Kang}. This situation is quite unsatisfactory since the 
charged Reissner-Nordstrom black hole makes its appearance in various modern
theories such as supergravity and superstring theories.
The aim of this paper is to show that it $\it{is}$ indeed possible to have 
the Reissner-Nordstrom black hole  as a solution in gravity localized models,
which include the Randall-Sundrum model in case of $AdS_5$.
To do that, we shall introduce the bulk $U(1)$ gauge field in the models,
which is needed to obtain Einstein's equations with energy-momentum tensor
associated with the electric field.

At this stage an interesting question arises with respect to the localization
of the bulk gauge field on a 3-brane in the Randall-Sundrum model
\cite{Randall2}.
In the Randall-Sundrum model, we are now familiar with the following fact
about 
the localization of bulk gauge field on the 3-brane by means of a 
gravitational interaction: zero mode of spin 1 bulk vector field is not 
localized neither on a brane with positive tension nor on a brane with
negative tension
\cite{Pomarol, Bajc}. Recently, this localization mechanism has been also
investigated 
in higher dimensional topological defect models \cite{Gregory, Vilenkin,
Gherghetta, 
Oda3, Gherghetta2} by the present author \cite{Oda3, Oda4}. 
In the analysis \cite{Oda3, Oda4}, we have clarified that zero mode of 
spin 1 vector field $\it{is}$ localized on a brane with positive tension only 
in a string-like defect model with codimension 2, but the other topological
defects 
cannot support zero mode of vector field on a 3-brane like the domain wall 
in the Randall-Sundrum model.

Accordingly, at first sight we might conclude that although we could
succeed in 
constructing the charged Reissner-Nordstrom black hole on a 3-brane by
including the 
bulk gauge field in the gravity localized models, the black hole, or more
precisely,
the charge carried by the black hole, cannot stay in our universe and enters
in the higher dimensional space-time except the string-like defect model.
Surprisingly enough, however, the charged Reissner-Nordstrom black hole
$\it{can}$ 
reside in our universe. Why is the charged black hole bound on 3-brane by
circumventing
the aforementioned no-go theorem? 
A key word here is 'zero mode'. Although it is certainly true that the
$\it{constant}$ 
zero mode of the bulk gauge field is not trapped on a 3-brane, the
$\it{non-constant}$ 
zero mode has a possibility of being trapped on the 3-brane. Indeed, this
phenomenon 
precisely occurs in the present situation as will seen later.

This paper is organized as follows. In the next section, we present our
setup and
derive the equations of motion from an action in a general space-time
dimension. 
In Section 3, we show that there are the Reissner-Nordstrom black holes as
solutions
to the equations of motion obtained in Section 2. Then, in Section 4, we
explain
why such charged black holes are localized on a 3-brane. 
The final section is devoted to discussions.

%%%%%%%%%%%%%%%%%%%%%%%%%%%%%%%%%%%%%%%%%%%%%%%%%%%%%%%%%%%%%%%%%%%%%
%%%%%%%%%%%%%%%%%%%%%%%%%%%%%%   SEC  2    %%%%%%%%%%%%%%%%%%%%%%%%%%
%%%%%%%%%%%%%%%%%%%%%%%%%%%%%%%%%%%%%%%%%%%%%%%%%%%%%%%%%%%%%%%%%%%%
\section{Setup}

We would like to consider $\it{gobal}$ topological defects in the extra
internal dimensions, which are described by a scalar field with an
orthogonal symmetry with the Higgs potential. In the defects, the scalar 
field takes the vanishing configuration at the core, whereas it has
the 'hedgehog' configuration outside the core. In this article, we
pay our attention to only the exterior geometry outside the core of the 
defects.

The action with which we start is that of gravity
in general $D$ dimensions, with the conventional Einstein-Hilbert
action, a cosmological constant $\Lambda$ and some matter action $S_m$:
%**   1 %%%%%%%%%%%%%%%%%%%%%%%%%%%%%%%%%%%%%%%%%%%%%%%%%%%%%%%%%
\begin{eqnarray}
S = \frac{1}{4 \kappa_D^2} \int d^D x  
\sqrt{-g} \left(R - 2 \Lambda \right) + S_m,
\label{1}
\end{eqnarray}
%%%%%%%%%%%%%%%%%%%%%%%%%%%%%%%%%%%%%%%%%%%%%%%%%%%%%%%%%%%%%%%%%%%
where $\kappa_D$ denotes the $D$-dimensional gravitational
constant with a relation $\kappa_D^2 = 8 \pi G_N = \frac{8 \pi}
{M_*^{D-2}}$ with $G_N$ and $M_*$ being the $D$-dimensional Newton 
constant and the $D$-dimensional Planck mass scale, respectively.
For simplicity, we set $\kappa_D$ to 1 from now on.
In this article, as $S_m$ we shall take the $U(1)$ Maxwell's action 
$S_A$ plus the $O(n)$ globally symmetric scalar action $S_{\Phi}$ 
with the Higgs potential ($S_m = S_A + S_{\Phi}$), which are defined as
%**   2 %%%%%%%%%%%%%%%%%%%%%%%%%%%%%%%%%%%%%%%%%%%%%%%%%%%%%%%%%
\begin{eqnarray}
S_A &=& - \frac{1}{4} \int d^D x \sqrt{-g} g^{MN} g^{RS} F_{MR} F_{NS}, \nn\\
S_{\Phi} &=& \int d^D x \sqrt{-g} \left\{ - \frac{1}{2} g^{MN} \partial_M
\Phi^a
\partial_N \Phi^a + \frac{\lambda}{4}(\Phi^a\Phi^a - \eta^2)^2 \right\},
\label{2}
\end{eqnarray}
%%%%%%%%%%%%%%%%%%%%%%%%%%%%%%%%%%%%%%%%%%%%%%%%%%%%%%%%%%%%%%%%%%%
where $F_{MN} = \partial_M A_N - \partial_N A_M$.
And we use the index convention such that 
$M, N, ...$ denote $D$-dimensional space-time indices, 
$\mu, \nu, ...$ $p$-dimensional brane ones, and $a, b, ...$
$n$-dimensional extra spatial ones, so the equality $D=p+n$
holds. 
Throughout this article we follow the standard 
conventions and notations of the textbook of Misner, Thorne and 
Wheeler \cite{Misner}. 

Taking a variation of the action (\ref{1}) with respect to the 
$D$-dimensional metric tensor $g_{MN}$ leads to Einstein's equations
%**   3 %%%%%%%%%%%%%%%%%%%%%%%%%%%%%%%%%%%%%%%%%%%%%%%%%%%%%%%%%
\begin{eqnarray}
R_{MN} - \frac{1}{2} g_{MN} R 
= - \Lambda g_{MN}  + 2 T_{MN},
\label{3}
\end{eqnarray}
%%%%%%%%%%%%%%%%%%%%%%%%%%%%%%%%%%%%%%%%%%%%%%%%%%%%%%%%%%%%%%%%%%%
where the energy-momentum tensor associated with Eq. (\ref{2})
is given by
%**   4 %%%%%%%%%%%%%%%%%%%%%%%%%%%%%%%%%%%%%%%%%%%%%%%%%%%%%%%%%
\begin{eqnarray}
T_{MN} &\equiv& T_{MN}(A) + T_{MN}(\Phi) \nn\\
&\equiv& - \frac{2}{\sqrt{-g}} \frac{\delta}{\delta g^{MN}} (S_A 
+ S_{\Phi}), \nn\\
T_{MN}(A) &=& F_{MP} F_N \ ^P - \frac{1}{4} \ g_{MN} F^2,  \nn\\
T_{MN}(\Phi) &=& \partial_M \Phi^a \partial_N \Phi^a  - \frac{1}{2} \ g_{MN} 
\partial_P \Phi^a \partial^P \Phi^a + \ g_{MN} \frac{\lambda}{4}
(\Phi^a\Phi^a - \eta^2)^2,
\label{4}
\end{eqnarray}
%%%%%%%%%%%%%%%%%%%%%%%%%%%%%%%%%%%%%%%%%%%%%%%%%%%%%%%%%%%%%%%%%%%
with the contraction over indices being achieved in terms of the
$D$-dimensional metric tensor $g^{MN}$.
The Bianchi identity gives rise to the conservation law for the 
energy-momentum tensor
%**   5 %%%%%%%%%%%%%%%%%%%%%%%%%%%%%%%%%%%%%%%%%%%%%%%%%%%%%%%%%
\begin{eqnarray}
\nabla^M T_{MN} = 0.
\label{5}
\end{eqnarray}
%%%%%%%%%%%%%%%%%%%%%%%%%%%%%%%%%%%%%%%%%%%%%%%%%%%%%%%%%%%%%%%%%%%
Finally, the equations of motion to vector fields $A_M$ read
%**   6 %%%%%%%%%%%%%%%%%%%%%%%%%%%%%%%%%%%%%%%%%%%%%%%%%%%%%%%%%
\begin{eqnarray}
\partial_M (\sqrt{-g} g^{M N} g^{R S} F_{NS}) = 0.
\label{6}
\end{eqnarray}
%%%%%%%%%%%%%%%%%%%%%%%%%%%%%%%%%%%%%%%%%%%%%%%%%%%%%%%%%%%%%%%%%%%

Next we shall propose our setup and ansatzs in order to solve the
equations of motion derived thus far. First of all, let us   
adopt the following cylindrical metric ansatz:
%**   7 %%%%%%%%%%%%%%%%%%%%%%%%%%%%%%%%%%%%%%%%%%%%%%%%%%%%%%%%%
\begin{eqnarray}
ds_D^2 &=& g_{MN} dx^M dx^N  \nn\\
&=& g_{\mu\nu}(x^M) dx^\mu dx^\nu + \tilde{g}_{ab}(x^c) dx^a dx^b  \nn\\
&=& e^{-A(r)} \hat{g}_{\mu\nu}(x^\lambda) dx^\mu dx^\nu + dr^2 
+ e^{-B(r)} d \Omega_{n-1}^2,
\label{7}
\end{eqnarray}
%%%%%%%%%%%%%%%%%%%%%%%%%%%%%%%%%%%%%%%%%%%%%%%%%%%%%%%%%%%%%%%%%%%
where d$\Omega_{n-1}^2$ stands for the metric on a unit 
$(n-1)$-sphere, which is concretely expressed in terms of the angular 
variables $\theta_i$ as
%**   8 %%%%%%%%%%%%%%%%%%%%%%%%%%%%%%%%%%%%%%%%%%%%%%%%%%%%%%%%%
\begin{eqnarray}
d \Omega_{n-1}^2 = d\theta_2^2 + \sin^2 \theta_2 d\theta_3^2 
+ \sin^2 \theta_2 \sin^2 \theta_3 d\theta_4^2 + \cdots
+ \prod_{i=2}^{n-1} \sin^2 \theta_i d\theta_n^2,
\label{8}
\end{eqnarray}
%%%%%%%%%%%%%%%%%%%%%%%%%%%%%%%%%%%%%%%%%%%%%%%%%%%%%%%%%%%%%%%%%%%
where each variable takes the range $0 \le r \le \infty$, $0 \le \theta_2 
\le 2 \pi$ and $0 \le \theta_i(i=3, 4, \cdots, n) \le \pi$, and the volume 
element is given by $\int d \Omega_{n-1} = \frac{2 \pi^{\frac{n}
{2}}}{\Gamma(\frac{n}{2})}$.

As mentioned above, global defects are regarded as taking 'hedgehog' 
configuration outside the core in the extra dimensions \cite{Vilenkin}
%**   9 %%%%%%%%%%%%%%%%%%%%%%%%%%%%%%%%%%%%%%%%%%%%%%%%%%%%%%%%%
\begin{eqnarray}
\Phi^a = \eta \frac{r^a}{r},
\label{9}
\end{eqnarray}
%%%%%%%%%%%%%%%%%%%%%%%%%%%%%%%%%%%%%%%%%%%%%%%%%%%%%%%%%%%%%%%%%%%
where $r^a$ is defined as 
%**   10 %%%%%%%%%%%%%%%%%%%%%%%%%%%%%%%%%%%%%%%%%%%%%%%%%%%%%%%%%
\begin{eqnarray}
r^a = (r\cos\theta_2, r\sin\theta_2\cos\theta_3, r\sin\theta_2
\sin\theta_3\cos\theta_4, \cdots, r \prod_{i=2}^{n-1} \sin\theta_i).
\label{10}
\end{eqnarray}
%%%%%%%%%%%%%%%%%%%%%%%%%%%%%%%%%%%%%%%%%%%%%%%%%%%%%%%%%%%%%%%%%%%
Then, the energy-momentum tensor $T^M_N(\Phi)$ takes the form
outside the defect core
%**   11 %%%%%%%%%%%%%%%%%%%%%%%%%%%%%%%%%%%%%%%%%%%%%%%%%%%%%%%%%
\begin{eqnarray}
T^r_r(\Phi) &=& -\frac{1}{2} (n-1) \eta^2 e^{B(r)}, \nn\\
T^{\theta_i}_{\theta_j}(\Phi) &=& -\frac{1}{2} (n-3) \eta^2 
e^{B(r)} \delta^i_j, \nn\\
T^\mu_\nu(\Phi) &=& -\frac{1}{2} (n-1) \eta^2 e^{B(r)} 
\delta^\mu_\nu.
\label{11}
\end{eqnarray}
%%%%%%%%%%%%%%%%%%%%%%%%%%%%%%%%%%%%%%%%%%%%%%%%%%%%%%%%%%%%%%%%%%%

{}For the field strength $F_{MN}$ of the $U(1)$ gauge field $A_M$, we take an
ansatz such that only the nonvanishing components are
%**   12 %%%%%%%%%%%%%%%%%%%%%%%%%%%%%%%%%%%%%%%%%%%%%%%%%%%%%%%%%
\begin{eqnarray}
F_{t \rho} = - F_{\rho t} = e^{-\frac{1}{2} A(r)} f_{t \rho}
= - e^{-\frac{1}{2} A(r)} f_{\rho t},
\label{12}
\end{eqnarray}
%%%%%%%%%%%%%%%%%%%%%%%%%%%%%%%%%%%%%%%%%%%%%%%%%%%%%%%%%%%%%%%%%%%
where we have denoted the coordinates $x^0$ and $x^1$ on a $(p-1)$-brane
as $t$ and $\rho$, respectively, and we assume that $f_{t \rho}$ is
a function depending on only the radial coordinate $\rho$ on a brane. 
Here we have
inserted the nontrivial $r$-dependent factor $e^{-\frac{1}{2} A(r)}$
in front of $f_{t \rho}$ for the purpose of canceling out the
$r$-dependent factor appearing in $T_{MN}(A)$ as will be seen shortly. 

With these ansatzs, after a straightforward but a little tedious
calculation, Einstein's equations (\ref{3}) reduce to 
%**   13 %%%%%%%%%%%%%%%%%%%%%%%%%%%%%%%%%%%%%%%%%%%%%%%%%%%%%%%%%
\begin{eqnarray}
e^A \hat{R} - \frac{p(n-1)}{2} A' B' - \frac{p(p-1)}{4} (A')^2
- \frac{(n-1)(n-2)}{4} (B')^2 \nn\\
+ (n-1)(n-2-2\eta^2) e^B - 2\Lambda = 0,
\label{13}
\end{eqnarray}
%%%%%%%%%%%%%%%%%%%%%%%%%%%%%%%%%%%%%%%%%%%%%%%%%%%%%%%%%%%%%%%%%%%
%**   14 %%%%%%%%%%%%%%%%%%%%%%%%%%%%%%%%%%%%%%%%%%%%%%%%%%%%%%%%%
\begin{eqnarray}
e^A \hat{R} + (n-2) B'' - \frac{p(n-2)}{2} A' B' 
- \frac{(n-1)(n-2)}{4} (B')^2  \nn\\
+ (n-3)(n-2-2\eta^2) e^B + p A'' -  \frac{p(p+1)}{4} (A')^2 - 2\Lambda = 0,
\label{14}
\end{eqnarray}
%%%%%%%%%%%%%%%%%%%%%%%%%%%%%%%%%%%%%%%%%%%%%%%%%%%%%%%%%%%%%%%%%%%
%**   15 %%%%%%%%%%%%%%%%%%%%%%%%%%%%%%%%%%%%%%%%%%%%%%%%%%%%%%%%%
\begin{eqnarray}
\hat{R}_{\mu\nu} - \frac{1}{2} \hat{g}_{\mu\nu} \hat{R} 
= - \Lambda_p \hat{g}_{\mu\nu} + 2 \hat{t}_{\mu\nu}.
\label{15}
\end{eqnarray}
%%%%%%%%%%%%%%%%%%%%%%%%%%%%%%%%%%%%%%%%%%%%%%%%%%%%%%%%%%%%%%%%%%%
In the above the prime denotes the differentiation with respect to $r$,
and then $\hat{R}_{\mu\nu}$ and $\hat{R}$ are respectively the Ricci tensor 
and the scalar curvature associated with the brane metric $\hat{g}_{\mu\nu}$.
Here we have defined the effective cosmological constant on the 
$(p-1)$-brane, $\Lambda_p$, by the equation
%**   16 %%%%%%%%%%%%%%%%%%%%%%%%%%%%%%%%%%%%%%%%%%%%%%%%%%%%%%%%%
\begin{eqnarray}
- e^{A} \Lambda_p = \frac{p-1}{2} (A'' - \frac{n-1}{2} A' B')
- \frac{p(p-1)}{8} (A')^2 \nn\\
+ \frac{n-1}{2} \left\{B'' - \frac{n}{4} (B')^2 + (n-2-2\eta^2) e^B 
\right\} - \Lambda.
\label{16}
\end{eqnarray}
%%%%%%%%%%%%%%%%%%%%%%%%%%%%%%%%%%%%%%%%%%%%%%%%%%%%%%%%%%%%%%%%%%%
In addition, the nonvanishing energy-momentum tensor $T_{\mu\nu}(A)$ 
takes the form
%**   17 %%%%%%%%%%%%%%%%%%%%%%%%%%%%%%%%%%%%%%%%%%%%%%%%%%%%%%%%%
\begin{eqnarray}
T_{\mu\nu}(A) = f_{\mu\lambda} f_\mu \ ^\lambda - \frac{1}{4} 
\hat{g}_{\mu\nu} f^2 \equiv \hat{t}_{\mu\nu},
\label{17}
\end{eqnarray}
%%%%%%%%%%%%%%%%%%%%%%%%%%%%%%%%%%%%%%%%%%%%%%%%%%%%%%%%%%%%%%%%%%%
where the indices are raised by means of $\hat{g}^{\mu\nu}$.
Note that as promised this energy-momentum tensor has no dependency 
on $A(r)$ due to Eq. (\ref{12}).

Furthermore, the conservation law for this energy-momentum tensor
(\ref{5}) reads
%**   18 %%%%%%%%%%%%%%%%%%%%%%%%%%%%%%%%%%%%%%%%%%%%%%%%%%%%%%%%%
\begin{eqnarray}
\hat{\nabla}_\mu \hat{t}^\mu \ _\nu = 0, \ A' \hat{t}^\mu \ _\mu 
- (n-1) \eta^2 B' e^B = 0.
\label{18}
\end{eqnarray}
%%%%%%%%%%%%%%%%%%%%%%%%%%%%%%%%%%%%%%%%%%%%%%%%%%%%%%%%%%%%%%%%%%%
Finally, Maxwell's equations (\ref{6}) become 
%**   19 %%%%%%%%%%%%%%%%%%%%%%%%%%%%%%%%%%%%%%%%%%%%%%%%%%%%%%%%%
\begin{eqnarray}
\partial_\mu (\sqrt{-\hat{g}} \hat{g}^{\mu\nu} \hat{g}^{\lambda\sigma} 
f_{\nu\sigma}) = 0.
\label{19}
\end{eqnarray}
%%%%%%%%%%%%%%%%%%%%%%%%%%%%%%%%%%%%%%%%%%%%%%%%%%%%%%%%%%%%%%%%%%%
Note that the raising and the lowering of the indices are respectively
carried out by the brane metric tensor $\hat{g}^{\mu\nu}$ and 
$\hat{g}_{\mu\nu}$ in all the reduced equations of motion, which is
crucial for the existence of a black hole solution on a brane.

%%%%%%%%%%%%%%%%%%%%%%%%%%%%%%%%%%%%%%%%%%%%%%%%%%%%%%%%%%%%%%%%%%%%%
%%%%%%%%%%%%%%%%%%%%%%%%%%%%%%   SEC  3    %%%%%%%%%%%%%%%%%%%%%%%%%%
%%%%%%%%%%%%%%%%%%%%%%%%%%%%%%%%%%%%%%%%%%%%%%%%%%%%%%%%%%%%%%%%%%%%%
\section{The Reissner-Nordstrom black hole solution}

We are now in a position to find the Reissner-Nordstrom black hole
on a $(p-1)$-brane in a warped geometry as a solution to a set of 
equations obtained so far.

Henceforth, let us confine ourselves to $p=4$, in other words,
a 3-brane, from a physical interest. (The generalization to an arbitrary
$p$ is straightforward.)  In addition to it, we require an ansatz
$\hat{R}=0$. The physical plausibility of this ansatz arises from 
the reasoning that
the scalar curvature of the Reissner-Nordstrom black hole is vanishing
although the Ricci tensor is not zero but proportional to stress-energy
tensor of the electric field. 
In what follows, we shall solve a set of equations for the domain wall $n=1$
first, and then for the higher dimensional defects $n \ge 2$.

\subsection{Domain wall (n=1) in five dimensions (D=5)}

In this subsection, we solve a set of equations in the case of
$n=1$ and five space-time dimensions. 

In this case, under the ansatzs $p=4$ and $\hat{R}=0$, the (rr)-component of
Einstein equations, (\ref{13}), reduces to the form
%**   20 %%%%%%%%%%%%%%%%%%%%%%%%%%%%%%%%%%%%%%%%%%%%%%%%%%%%%%%%%
\begin{eqnarray}
 - 3 (A')^2 - 2 \Lambda  = 0,
\label{20}
\end{eqnarray}
%%%%%%%%%%%%%%%%%%%%%%%%%%%%%%%%%%%%%%%%%%%%%%%%%%%%%%%%%%%%%%%%%%%
whose solution is given by
%**   121%%%%%%%%%%%%%%%%%%%%%%%%%%%%%%%%%%%%%%%%%%%%%%%%%%%%%%%%%
\begin{eqnarray}
A =  c r,
\label{21}
\end{eqnarray}
%%%%%%%%%%%%%%%%%%%%%%%%%%%%%%%%%%%%%%%%%%%%%%%%%%%%%%%%%%%%%%%%%%%
with the definition of a constant $c \equiv \sqrt{\frac{-2 \Lambda}{3}}$. 
Here three remarks are in order. The first is that the positivity of
$(A')^2$ requires the negativity of a bulk cosmological constant,
i.e., anti-de Sitter space as in the Randall-Sundrum model.
Actually, when the brane geometry is completely flat ($\hat{g}_{\mu\nu}
= \eta_{\mu\nu}$) and chargeless,
the present model becomes equivalent to (a half $AdS_5$ slice of)
the Randall-Sundrum model
where a single 3-brane sits at the origin $r=0$ and a fifth dimension
is noncompact. The second remark is related to the sign in front of
the positive constant $c$ where we have selected a positive sign to
bind the graviton on our 3-brane. The final remark is the integration
constant in the right hand side of (\ref{21}). It is obvious that the
constant can be set to zero by the redefinition of the brane coordinates
$x^\mu$.

Then substituting (\ref{21}) into (\ref{16}), we find that the effective
cosmological constant on the brane vanishes, $\Lambda_4=0$. Next, under
$\Lambda_4=0$ and the ansatz $\hat{R}=0$, we see from (\ref{15}) that   
%**   22 %%%%%%%%%%%%%%%%%%%%%%%%%%%%%%%%%%%%%%%%%%%%%%%%%%%%%%%%%
\begin{eqnarray}
\hat{R}_{\mu\nu} =  2 \hat{t}_{\mu\nu}.
\label{22}
\end{eqnarray}
%%%%%%%%%%%%%%%%%%%%%%%%%%%%%%%%%%%%%%%%%%%%%%%%%%%%%%%%%%%%%%%%%%%
Indeed, this equation is consistent with our ansatz $\hat{R}=0$
from Eq. (\ref{18}).

In consequence, the remaining equations which we should solve are 
Eqs. (\ref{22}), (\ref{18}) and (\ref{19}). (Note that Eq. (\ref{14})
stems from the $(\theta_i\theta_i)$-component of Einstein's equations
so this equation does not exist in the case of domain wall.)
Provided that we
choose $f_{t\rho}=\frac{Q}{\rho^2}$ with $Q$ being a constant
corresponding to the electric charge, as a special solution
these equations give us the electrically charged Reissner-Nordstrom 
black hole geometry on the brane \cite{Misner}. 
Accordingly, we arrive at the following line element as desired
%**   23 %%%%%%%%%%%%%%%%%%%%%%%%%%%%%%%%%%%%%%%%%%%%%%%%%%%%%%%%%
\begin{eqnarray}
ds_5^2 = e^{- c r} d\hat{s}^2 + dr^2,
\label{23}
\end{eqnarray}
%%%%%%%%%%%%%%%%%%%%%%%%%%%%%%%%%%%%%%%%%%%%%%%%%%%%%%%%%%%%%%%%%%%
where $d\hat{s}^2$ is the line element of the Reissner-Nordstrom black 
hole geometry on the brane
%**   24 %%%%%%%%%%%%%%%%%%%%%%%%%%%%%%%%%%%%%%%%%%%%%%%%%%%%%%%%%
\begin{eqnarray}
d\hat{s}^2 &=& \hat{g}_{\mu\nu}(x^\lambda) dx^\mu dx^\nu, \nn\\
&=& -(1 - \frac{2M}{\rho} + \frac{Q^2}{\rho^2}) dt^2 
+ \frac{1}{1 - \frac{2M}{\rho} + \frac{Q^2}{\rho^2}} d\rho^2
+ \rho^2 (d\phi^2 + \sin^2\phi d\varphi^2).
\label{24}
\end{eqnarray}
%%%%%%%%%%%%%%%%%%%%%%%%%%%%%%%%%%%%%%%%%%%%%%%%%%%%%%%%%%%%%%%%%%%

\subsection{Higher dimensional defects (n $\ge$ 2) in D dimensions 
(D = 4 + n $\ge$ 6)}

Now we turn our attention to the more general defect with codimension
$n$ in $D$ space-time dimensions. A defect with codimension 2 in six 
space-time dimensions is called a string-like defect. Similarly, a defect
with codimension 3 and 4 in seven and eight space-time dimensions is called
a monopole-like and an instanton-like defect, respectively. Of course, all
the defects are a 3-brane. (Recall that we have set $p=4$.)
In this case, even if we take the ansatzs $p=4$ and $\hat{R}=0$ like
the domain wall, Einstein equations (\ref{13}) and (\ref{14}) do not
reduce to a simple and manageable expression owing to the existence
of the nontrivial term $e^B$, which comes from the fact that in the
$n$ extra dimensions with $n \ge 2$ the space is not flat but curved.
(Note that the extra space with $n=2$ is marginal since the space is
conformally flat, so there is a nontrivial solution in this case,
which was discovered by Gregory \cite{Gregory}.)   

An important observation here is that one must set
a function $B(r)$ to a constant because otherwise we encounter a 'naked'
singularity at somewhere $r$. This observation can be checked by calculating
the bulk scalar curvature and higher orders of the curvature tensor
\cite{Gherghetta2}. Thus, we assume the form
%**   25 %%%%%%%%%%%%%%%%%%%%%%%%%%%%%%%%%%%%%%%%%%%%%%%%%%%%%%%%%
\begin{eqnarray}
B = - 2 \ln R_0,
\label{25}
\end{eqnarray}
%%%%%%%%%%%%%%%%%%%%%%%%%%%%%%%%%%%%%%%%%%%%%%%%%%%%%%%%%%%%%%%%%%%
where we selected the constant to be a specific form for later
convenience.

With the ansatzs $p=4$, $\hat{R}$ and (\ref{25}), Einstein equations 
(\ref{13}) and (\ref{14}) read respectively
%**   26 %%%%%%%%%%%%%%%%%%%%%%%%%%%%%%%%%%%%%%%%%%%%%%%%%%%%%%%%%
\begin{eqnarray}
- 3 (A')^2 + (n-1)(n-2-2\eta^2) R_0^{-2} - 2 \Lambda  = 0,
\label{26}
\end{eqnarray}
%%%%%%%%%%%%%%%%%%%%%%%%%%%%%%%%%%%%%%%%%%%%%%%%%%%%%%%%%%%%%%%%%%%
%**   27 %%%%%%%%%%%%%%%%%%%%%%%%%%%%%%%%%%%%%%%%%%%%%%%%%%%%%%%%%
\begin{eqnarray}
4 A'' - 5 (A')^2 + (n-3)(n-2-2\eta^2) R_0^{-2}- 2 \Lambda  = 0.
\label{27}
\end{eqnarray}
%%%%%%%%%%%%%%%%%%%%%%%%%%%%%%%%%%%%%%%%%%%%%%%%%%%%%%%%%%%%%%%%%%%
It is straightforward to solve the equations (\ref{26}) and (\ref{27})
whose result is given by
%**   28 %%%%%%%%%%%%%%%%%%%%%%%%%%%%%%%%%%%%%%%%%%%%%%%%%%%%%%%%%
\begin{eqnarray}
A &=&  c r, \nn\\
c &=& \sqrt{\frac{-2 \Lambda}{n+2}}, \nn\\
R_0^2 &=& \frac{(n+2)(n-2-2\eta^2)}{2 \Lambda}.
\label{28}
\end{eqnarray}
%%%%%%%%%%%%%%%%%%%%%%%%%%%%%%%%%%%%%%%%%%%%%%%%%%%%%%%%%%%%%%%%%%%
This solution coincides with the one found in Ref. \cite{Vilenkin}. 

Note that this solution also has a warp factor and 
makes sense only in anti-de Sitter space.
Interestingly enough, in this case as well, the brane cosmological
constant is zero, so the same argument as in the domain wall gives us the 
Reissner-Nordstrom black hole solution on a 3-brane as a special
solution. Consequently, we have the bulk geometry with the 
Reissner-Nordstrom black hole as the brane geometry
%**   29 %%%%%%%%%%%%%%%%%%%%%%%%%%%%%%%%%%%%%%%%%%%%%%%%%%%%%%%%%
\begin{eqnarray}
ds_D^2 = e^{-cr} d\hat{s}^2 + dr^2 + R_0^2 d \Omega_{n-1}^2,
\label{29}
\end{eqnarray}
%%%%%%%%%%%%%%%%%%%%%%%%%%%%%%%%%%%%%%%%%%%%%%%%%%%%%%%%%%%%%%%%%%%
where $d\hat{s}^2$, and the constants $c$ and $R_0$ are given by
Eqs. (\ref{24}) and (\ref{28}), respectively.

%%%%%%%%%%%%%%%%%%%%%%%%%%%%%%%%%%%%%%%%%%%%%%%%%%%%%%%%%%%%%%%%%%%%%
%%%%%%%%%%%%%%%%%%%%%%%%%%%%%%   SEC  4    %%%%%%%%%%%%%%%%%%%%%%%%%%
%%%%%%%%%%%%%%%%%%%%%%%%%%%%%%%%%%%%%%%%%%%%%%%%%%%%%%%%%%%%%%%%%%%%%
\section{Localization of Reissner-Nordstrom black hole on a 3-brane}

In the previous section, we have shown that the charged
Reissner-Nordstrom black hole exists on a 3-brane 
in the gravity localized models by introducing
the $U(1)$ bulk gauge field. But our knowledge of localization
of gauge field on the topological defects \cite{Pomarol, Bajc, Oda3,
Oda4} gives us a fear that the Reissner-Nordstrom black hole which
we have found may not be localized on the brane but live in the whole
space-time.
In what follows, we shall show that the Reissner-Nordstrom black hole
is in fact confined to the 3-brane.

As can be seen in Eqs. (\ref{23}) and (\ref{29}),
the 5D line element (\ref{23}) has a similar structure to the general
one (\ref{29}). As a result of this fact, it is known that in the both 
cases, zero mode of the bulk gauge field is $\it{not}$ localized on a 3-brane
\cite{Pomarol, Bajc, Oda3, Oda4}.
To save space, we shall explicitly prove that the Reissner-Nordstrom 
black hole which we have found in the previous section is in fact 
confined to a 3-brane only for the solution (\ref{29}). 
The case of 5D solution (\ref{23}) is a direct
consequence of the proof. 

Let us start with the action of $U(1)$ vector field:
%**   30 %%%%%%%%%%%%%%%%%%%%%%%%%%%%%%%%%%%%%%%%%%%%%%%%%%%%%%%%%
\begin{eqnarray}
S_A = - \frac{1}{4} \int d^D x \sqrt{-g} g^{M N} g^{R S}
F_{MR} F_{NS}.
\label{30}
\end{eqnarray}
%%%%%%%%%%%%%%%%%%%%%%%%%%%%%%%%%%%%%%%%%%%%%%%%%%%%%%%%%%%%%%%%%%%
{}From (\ref{29}) with the help of (\ref{8}), we can calculate 
%**   31 %%%%%%%%%%%%%%%%%%%%%%%%%%%%%%%%%%%%%%%%%%%%%%%%%%%%%%%%%
\begin{eqnarray}
\sqrt{-g} = R_0^{n-1} e^{-2A} (\sin\theta_2)^{n-2} (\sin\theta_3)^{n-3}
\cdots (\sin\theta_{n-2})^2 (\sin\theta_{n-1}) \sqrt{-\hat{g}}.
\label{31}
\end{eqnarray}
%%%%%%%%%%%%%%%%%%%%%%%%%%%%%%%%%%%%%%%%%%%%%%%%%%%%%%%%%%%%%%%%%%%
Then, using Eqs. (\ref{12}), (\ref{29}) and (\ref{31}), we can rewrite
the action as follows:
%**   32 %%%%%%%%%%%%%%%%%%%%%%%%%%%%%%%%%%%%%%%%%%%%%%%%%%%%%%%%%
\begin{eqnarray}
S_A = - \frac{1}{4} \frac{2 \pi^{\frac{n}
{2}}}{\Gamma(\frac{n}{2})} R_0^{n-1} \int_0^{\infty} dr e^{-A}
\int d^4x \sqrt{-\hat{g}} \hat{g}^{\mu\nu} \hat{g}^{\lambda\sigma}
f_{\mu\lambda} f_{\nu\sigma} + \cdots.
\label{32}
\end{eqnarray}
%%%%%%%%%%%%%%%%%%%%%%%%%%%%%%%%%%%%%%%%%%%%%%%%%%%%%%%%%%%%%%%%%%%

The existence of a normalizable mode of gauge field on a 3-brane
requires that the coefficient in front of 4D gauge action should be
finite \cite{Bajc, Oda3}. Therefore, provided that we define 
the nontrivial coefficient in (\ref{32}) as $I_1$, the requirement that
the Reissner-Nordstrom black hole is confined to a 3-brane
is that $I_1$ is finite. This is indeed the case as follows:
%**   33 %%%%%%%%%%%%%%%%%%%%%%%%%%%%%%%%%%%%%%%%%%%%%%%%%%%%%%%%%
\begin{eqnarray}
I_1 \equiv \int_0^{\infty} dr e^{-A} = \int_0^{\infty} dr e^{-c r}
= \frac{1}{c} < \infty.
\label{33}
\end{eqnarray}
%%%%%%%%%%%%%%%%%%%%%%%%%%%%%%%%%%%%%%%%%%%%%%%%%%%%%%%%%%%%%%%%%%%

It is quite of importance to consider why $I_1$ becomes finite in the
present situation since the corresponding coefficient in a Minkowski
flat 3-brane is known to be divergent so that the bulk gauge field cannot be
trapped on the brane by means of only a gravitational interaction
\cite{Bajc, Oda3}.
Note that an exponential factor $e^{-A}$ in (\ref{33}) comes from (\ref{12}). 
(Recall that as mentioned in Section 1, 
if we do not have this exponential damping factor as in the flat brane, 
$I_1$ becomes divergent, leading to the conclusion that zero mode of 
spin 1 bulk vector field is not localized neither on a brane with positive 
tension nor on a brane with negative tension \cite{Pomarol, Bajc}.)
This factor was introduced in order to have the Reissner-Nordstrom black 
hole as a solution to Einstein's equations. 
(See Eqs. (\ref{15}), (\ref{17}) and recall an ansatz 
$f_{t\rho} = \frac{Q}{\rho^2}$.) Accordingly, the existence of the
Reissner-Nordstrom black hole solution naturally causes the localization
of the black hole charge on a 3-brane! We believe that this is one of striking
phenomena in the gravity localized models. Put differently,
a no-go theorem with regard to the non-localization of gauge field 
\cite{Pomarol, Bajc} holds only when its zero mode is a constant mode
with respect to the radial coordinate $r$, whereas in the case at hand
zero mode explicitly depends on the coordinate $r$ so this no-go theorem 
is not valid now.

%%%%%%%%%%%%%%%%%%%%%%%%%%%%%%%%%%%%%%%%%%%%%%%%%%%%%%%%%%%%%%%%%%%%%
%%%%%%%%%%%%%%%%%%%%%%%%%%%%%%   SEC  5    %%%%%%%%%%%%%%%%%%%%%%%%%%
%%%%%%%%%%%%%%%%%%%%%%%%%%%%%%%%%%%%%%%%%%%%%%%%%%%%%%%%%%%%%%%%%%%%%
\section{Discussions}

In this article we have investigated the possibility of having
the electrically charged Reissner-Nordstrom black hole in the gravity
localized models in a general dimension. To do so, it was necessary
to introduce the bulk gauge field in order to make Einstein's equations
with the stress-energy tensor of the electric field. But the story was not
so simple since it is well known that the bulk gauge field is not
generally localized on a 3-brane. Thus, even if we could construct the charged
Reissner-Nordstrom black hole on a 3-brane, the black hole, or more
precisely speaking its electric charge, may not be trapped
on the brane and propagate freely in the whole space-time.
In other words, the charged black hole originally living in our world
may lose its charge into the extra space and become the Schwarzschild
black hole on a 3-brane as time passes by. 

One of striking features is that the charged Reissner-Nordstrom black hole 
in the gravity localized models is indeed localized on our universe and the
localization mechanism is provided by Einstein's equations in a 
remarkable way.

In a flat Minkowski $(p-1)$-brane, spin 1/2 and 
3/2 fermionic fields are known to be localized on the brane with the 
exponentially rising warp factor \cite{Grossman, Bajc, Oda3}. 
We wish to know in future whether or not
a similar phenomenon to the trapping of the charged black hole could happen 
in a physical situation relevant to those fermionic fields. 

\vs 1
%%%%%%%%%%%%%%%%%%%%%%%%%%%%%%%%%%%%%%%%%%%%%%%%%%%%%%%%%%%%%%%%%%
%%%%%%%%%%%%%%%%%%%%%%%% reference %%%%%%%%%%%%%%%%%%%%%%%%%%%%%%%
%%%%%%%%%%%%%%%%%%%%%%%%%%%%%%%%%%%%%%%%%%%%%%%%%%%%%%%%%%%%%%%%%%

\end{document}